\documentclass[aps,pre,superscriptaddress,twocolumn,amsmath,amssymb,showpacs]{revtex4-1}
\usepackage{amsmath}
\usepackage{amssymb}
\usepackage{graphicx}
\usepackage{float}
\usepackage{subfigure}
\usepackage{color}
\usepackage{comment}
\usepackage[utf8x]{inputenc}
\usepackage{tabularx}
\usepackage{array}

\begin{document}
\title{Semianalytical solutions of Ising-like and Potts-like magnetic polymers on the Bethe lattice}
\author{Nathann T. Rodrigues}
\email{nathan.rodrigues@ufv.br}
\affiliation{Instituto de Física, Universidade Federal Fluminense, Avenida Litor\^anea s/n, 24210-346 Niter\'oi, Rio de Janeiro, Brazil}
\affiliation{Departamento de Física, Universidade Federal de Viçosa, 36570-900, Vi\c cosa, MG, Brazil}
\author{Tiago J. Oliveira}
\email{tiago@ufv.br}
\affiliation{Departamento de Física, Universidade Federal de Viçosa, 36570-900, Vi\c cosa, MG, Brazil}
\date{\today}

\begin{abstract}
We study magnetic polymers, defined as self-avoiding walks where each monomer $i$ carries a ``spin'' $s_i$ and interacts with its first neighbor monomers, let us say $j$, via a coupling constant $J(s_i,s_j)$. Ising-like [$s_i = \pm 1$, with $J(s_i,s_j) = \varepsilon s_i s_j$] and Potts-like [$s_i = 1,\ldots,q$, with $J(s_i,s_j)=\varepsilon_{s_i} \delta(s_i,s_j)$] models are investigated. Some particular cases of these systems have recently been studied in the continuum and on regular lattices, and are related to interesting applications. Here, we solve these models on Bethe lattices of ramification $\sigma$, focusing on the ferromagnetic case in zero external magnetic field. In most cases, the phase diagrams present a non-polymerized (NP) and two polymerized phases: a paramagnetic (PP) and a ferromagnetic (FP) one. However, quite different thermodynamic properties are found depending on $q$ in the Potts-like polymers and on whether one uses the Ising or Potts coupling in the two-state systems. For instance, for the standard Potts model (where $\varepsilon_1=\cdots=\varepsilon_q$) with $q=2$, beyond the $\theta$-point (where the critical and discontinuous NP-PP transition lines meet), a second tricritical point connecting a critical and a discontinuous transition line between the PP-FP phases is found in the system. A triple point where the NP-PP, NP-FP and PP-FP first-order transition lines meet is also present in the phase diagram. For $q \ge 3$ the PP-FP transition is always discontinuous, and the scenario with the triple and $\theta$ points appears for $q \le 6$. Interestingly, for $q \ge 7$, as well as for the Ising-like model the $\theta$ point becomes metastable and the critical NP-PP transition line ends at a critical-end-point (CEP), where it meets the NP-FP and PP-FP coexistence lines. Importantly, these results indicate that when $q\le 6$ the spin ordering transition is preceded by the polymer collapse transition, whereas for $q\ge 7$ and in the Ising case these transitions happen together at the CEPs. Some interesting non-standard Potts models are also studied, such as the lattice version of the model for epigenetic marks in the chromatin introduced in [PRX {\bf 6}, 041047 (2016)]. In addition, the solution of the dilute Ising and dilute Potts models on the Bethe lattice are also presented here, once they are important to understand the PP-FP transitions. 
\end{abstract}

\maketitle

\section{Introduction}
\label{secIntro}

Simple model systems have played a major role in the development of the modern theory of equilibrium phase transitions and critical phenomena \cite{StanleyBook,YeomansBook,StanleyRevModPhys}. Beyond usually providing deep insight on the real systems they are intended to describe, the study of such models is justified also by the universality observed near criticality. In this context, the most paradigmatic example is certainly the Ising model \cite{Ising25}. Introduced as a model for an anisotropic magnet --- with ``spins" $s_i = \pm 1$ placed on the sites of a given lattice, where each pair $(i,j)$ of nearest neighbor (NN) spins has an energy $-\varepsilon s_i s_j$ --- its critical properties are related to those of order-disorder transitions in binary alloys, gas-liquid critical phenomena, phase separation in liquid mixtures and a variety of other examples. Several generalizations of the Ising model have appeared in the course of time, among which the Potts model \cite{Potts} is of particular interest here. In this model, each lattice site can be found in one of $q$ states and each pair of NN sites in the same state, contributes an energy $-\varepsilon$ to the system. This model is related to a number of experimental and model systems \cite{Wu}, being, e.g., mapped onto the Ising model for $q=2$. The three-state (related, e.g., to the famous Baxter's hard-hexagon model \cite{Baxter,BaxterBook}) and the four-state Potts universality classes have been also widely investigated. For $q \ge 5$ the order-disorder transition in the system is discontinuous \cite{Wu}.

Dilute polymers are another example of a very important system which can be modeled (in a coarse-grained approach) by simple lattice models \cite{desCloizeaux,Vanderzande}. In fact, the critical properties of polymer chains have been widely studied through self-avoiding walks (SAWs). By including an attractive interaction between NN monomers nonconsecutive along these walks, one obtains the interacting SAW (ISAW) model \cite{Orr}, which is a paradigmatic model for studying the polymer collapse transition. Due to the entropy-energy dispute, at high temperatures (and/or in a good solvent) the polymer stays in a coil configuration, whereas at low temperatures (and/or in a poor solvent) it assumes a globular shape \cite{Flory1,Flory2}. At a $\theta$ temperature (and/or in a $\theta$ solvent) the system undergoes a coil-globule transition and, as demonstrated by De Gennes \cite{deGennes1,deGennes2}, the $\theta$-point is a tricritical point, with the coil phase being a critical condition. Alternatives to the ISAW model, considering on-site interactions, may also display a continuous collapse transition, examples including the vertex ISAW (VISAW) model \cite{VISAW} (whose critical exponents in two-dimensions are different from those for the ISAW model \cite{ISAWexp}), interacting self-avoiding trails \cite{ISAT} (whose nature and universality class of the collapse transition in two-dimensions is still a subject of debate \cite{ISATuc}), among others \cite{WuBradley,MMS}. Modified ISAW models, considering, e.g., interactions up to second neighbors \cite{ISAW2NN} or bond-bond interactions \cite{Machado} have been also analyzed.

Here, we are interested in another type of polymer model, where each monomer of a SAW carries a ``spin" and interacts with its NN monomers (including the bonded ones) through a magnetic coupling. Three versions of this system have been considered in the literature: \textit{i}) the case where dynamic spins are placed on static SAWs \cite{Chakrabarti,*Aerstens,*Papale}; \textit{ii}) the opposite situation, where the spin configuration is ``frozen" along a dynamic chain \cite{Archontis}; and \textit{iii}) the most general and interesting case, where both the spins and polymer conformations may change, giving rise to \textit{magnetic polymers} \cite{Garel,Foster,Faizullina}. We remark that such polymers are very appealing, once flexible magnetic materials can be useful in a large number of applications \cite{Fink}. 

The ferromagnetic Ising-like magnetic polymers were firstly studied by Garel \textit{et al.} \cite{Garel}, via mean-field calculations and Monte Carlo (MC) simulations on the cubic lattice, where evidence for a first-order collapse transition, occurring concomitantly with the paramagnetic-to-ferromagnetic transition, was found for zero and small external magnetic field ($h$), whereas for large $h$ the continuous $\theta$ transition was recovered. In a recent work, Foster and Majumdar \cite{Foster} reported new MC simulations for this model confirming the first-order nature of the transition for $h=0$, but casting doubt on the scenario for $h \neq 0$. For the Ising-like model with $h=0$ on the square lattice, a continuous collapse transition (once again happening together with the magnetic transition) was found in Ref. \cite{Foster}. A similar result was independently obtained by Faizullina \textit{et al.} \cite{Faizullina}. Interestingly, these MC studies on the square lattice indicate that some critical exponents agree with the Ising ones while others are close to those for $\theta$ universality class \cite{Foster,Faizullina}. As an aside, we notice that a bond-fluctuating version of the Ising-like polymers has also been investigated by Luo and collaborators \cite{Luo}.

 In the case of Potts-like magnetic polymers, the Hamiltonian of the combined $n$-vector model and the $q$-state Potts model was proposed long ago (and studied in the limit of $n \rightarrow 0$ and $q \rightarrow 1$ via renormalization group calculations) as a model for vulcanisation \cite{Coniglio}. More recently, the $q=3$ case in the continuum space was used by Michieletto \textit{et al.} \cite{Michieletto} to investigate the effect of the spreading of epigenetic marks on the chromatin folding. By considering one of the three states as nonmagnetic (or unmarked) and attractive interactions between pairs of marked monomers with the same ``color", a first-order transition between an epigenetically disordered coil phase and an epigenetically coherent chromatin globule was found in \cite{Michieletto}.
 
 In this paper, we solve the Ising-like and Potts-like magnetic polymers on Bethe lattices with ramification $\sigma$, considering ferromagnetic interactions and zero external magnetic field. The (mean-field) grand-canonical phase diagrams present a series of interesting results, not anticipated in previous studies of these systems. To name a few, the $\theta$-point, a second tricritical point (related to the magnetic transition) and a triple point is found for the two-state Potts polymers, whereas in the Ising case the $\theta$-point becomes metastable, the triple point disappears and the collapse transition takes place (together with the magnetic transition) at a critical-end-point (CEP). A similar scenario is found for Potts systems with $q \ge 7$. For $q\le 6$, on the other hand, the spin ordering transition occurs separately of the polymer collapse transition. The thermodynamic properties of the lattice version of the model for epigenetic marks \cite{Michieletto}, of some other magnetic polymers with nonmagnetic monomers, as well as of the dilute Ising and dilute Potts models on the Bethe lattice are also investigated.

The outline of this paper is as follows. In Sec. \ref{secModel} we define the magnetic polymer models and devise their solution on the Bethe lattice. A comparison between the classical ISAW model and the $q=1$ Potts-like polymers is presented in Sec. \ref{secResq1}. Ising-like and $q=2$ Potts-like polymers are discussed in Sec. \ref{secResq2}, while results for the Potts-like models with $q \ge 3$ are presented \ref{secResqmi3}. Section \ref{secResNSPotts} brings results for some interesting Potts-like polymers with nonmagnetic monomers, including the lattice version of the model for epigenetic marks discussed above. Our final discussions and conclusions are summarized in Sec. \ref{secConc}. Some details on the exact calculation of the $\theta$-points are given in Appendix \ref{secApendTP}. Appendix \ref{secApPottsDil} presents the solutions of the dilute Ising and dilute Potts models on the BL.

\section{Models and their solution on the Bethe lattice}
\label{secModel}

\subsection{Models}

\begin{figure}[t]
 \includegraphics[width=8.3cm]{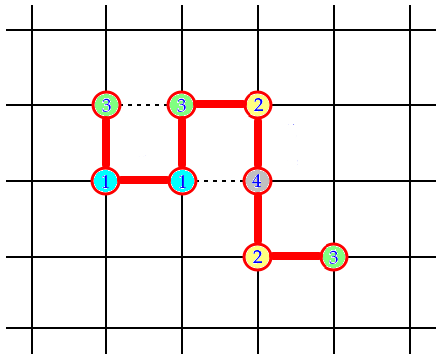}
 \caption{Illustration of a $q=4$ Potts-like polymer with eight monomers on the square lattice. The different numbers (and colors) denote the four possible states for each monomer. The thicker (red) lines represent the bonds connecting consecutive monomers in the walk. The dotted lines indicate the non-consecutive nearest neighbor monomers. The statistical weight of this configuration is $z^8 \omega_{1} \omega_{3}$.}
 \label{fig1}
\end{figure}

Polymers in dilute solutions are usually modelled as self-avoiding walks (SAWs), i.e., a sequence of $N$ monomers (placed on the sites of a given lattice of spacing $a$) connected by $N-1$ bonds (of length $a$, placed on the lattice edges) in a way that each site is visited by at most one monomer. See Fig. \ref{fig1}. Considering that each monomer $i$ carries a ``spin" (or ``color") $s_i=1,\ldots,q$ and interacts with a nearest neighbor (NN) monomer $j$ (including those consecutive along the chain) through a coupling constant $J(s_i,s_j)=\varepsilon_{s_i}\delta(s_i,s_j)$, where $\delta(s_i,s_j)$ is the Kronecker delta function, one obtains Potts-like magnetic polymers. The Ising-like model is given by $q=2$, $\varepsilon_1=\varepsilon_2$ and $J(s_i,s_j)=\varepsilon_{s_i} [2\delta(s_i,s_j)-1]$. Therefore, the grand-canonical energy of a given walk $\Omega_N$ reads
\begin{eqnarray}
\mathcal{H} _{\Omega_N} = - \sum_{\left<i,j\right>} J(s_i,s_j) - \mu N,
\end{eqnarray}
where $\mu$ is the chemical potential and the sum runs over all pairs of NN monomers, regardless of being bonded or not. Note that we are assuming here that there is no external magnetic field applied in the system. Moreover, following previous works on magnetic polymers on regular lattices \cite{Garel,Foster,Faizullina}, only the ferromagnetic case, corresponding to $\varepsilon_{s_i} \ge 0$, will be analyzed here. Hereafter, we will work with the monomer fugacity $z = e^{\mu/k_B T}$ and the weights $\omega_s = e^{\varepsilon_{s}/k_B T}$, so that the partition function of the Potts-like systems can be written as
\begin{eqnarray}
Y^{(P)} = \sum_{N} \sum_{\Omega_N} \sum_{\{s_1,\ldots,s_N\}} z^N \omega_{1}^{M_{1,1}} \omega_{2}^{M_{2,2}} \cdots \omega_{q}^{M_{q,q}},
\end{eqnarray}
where the second sum is carried over all SAW's of size $N$, while the third one runs over all spin configurations for a given SAW. The exponents $M_{j,j}(\Omega_N;\{s_1,\ldots,s_N\})$, for $j=1,\ldots,q$, denote the total number of pairs of NN monomers of type $(j,j)$. Although we are defining the model with a generalized Potts coupling, in most cases we will investigate the standard model, where $\omega_1 = \cdots = \omega_q = \omega$.

For the Ising-like model one has
\begin{eqnarray}
Y^{(I)} =  \sum_{N} \sum_{\Omega_N} \sum_{\{s_1,\ldots,s_N\}} z^N \omega^{M},
\end{eqnarray}
where $M = M_{1,1} + M_{2,2} - M_{1,2}$, with $M_{1,2}$ being the total number of NN monomers with different spins.

\subsection{Bethe lattice solution}

\begin{figure}[t]
 \includegraphics[width=8.cm]{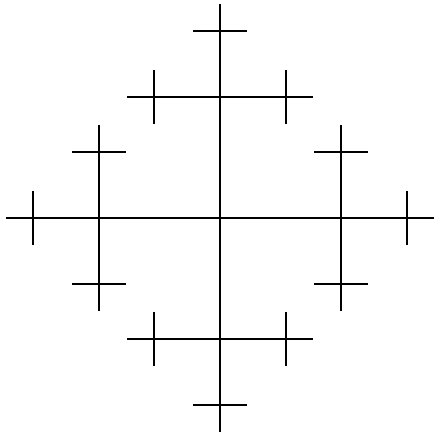}
 \caption{Symmetric Cayley tree with ramification $\sigma=3$ and three generations, starting from a central site.}
 \label{fig2}
\end{figure}

A Cayley tree is a connected graph, without loops, where all vertices (sites) have the same coordination number $\sigma+1$, with exception of the boundary ones, which have coordination 1. Hence, a symmetric Cayley tree can be built by connecting $\sigma+1$ edges to a central site and then successively adding $\sigma$ edges to each boundary site of the previous generation (see Fig. \ref{fig2}). The Bethe lattice (BL) is the core of the infinite Cayley tree \cite{BaxterBook}.

\begin{figure}[t]
 \includegraphics[width=8.5cm]{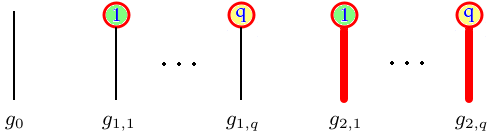}
 \caption{Possible configurations for the root sites for $q$-state Potts-like polymers on the BL. Following the notation in Fig. \ref{fig1}, different numbers (and colors) represent the different Potts states, while the thicker (red) lines denote polymer bonds. The leftmost configuration ($g_0$) corresponds to an empty root site.}
 \label{fig3}
\end{figure}

In order to solve a given model on the BL, we may write down recursion relations for partial partition functions (ppf's), considering the process of building a subtree with $M+1$ generations by connecting $\sigma$ subtrees, with $M$ generations each, to a root site of a rooted edge. As is shown in Fig. \ref{fig3}, this site can be empty (corresponding to ppf $g_0$), occupied by a monomer with spin $s$ without a bond in the root edge (ppf's $g_{1,s}$) or occupied by a monomer with spin $s$ with a bond in the rooted edge (ppf's $g_{2,s}$, with $s=1,\ldots,q$). So, there exist $2 q+1$ configurations for the root and related ppf's for the Potts-like polymers, and five of them in the Ising-like case. As it is always done in solutions of polymer models on the BL, we will consider that the chains start and end at the boundary of the tree. For later convenience, we will assign different fugacities $z_s$ for the monomers according to their spin $s$, but at the end we will always work with the same fugacity $z=z_1=\cdots=z_q$. In this way, \textit{for the Potts-like polymers} the recursion relations for the ppf's read:
\begin{subequations}
\label{eqPPFs}
\begin{eqnarray}
 g_0' = \left( g_0 + \sum_{i=1}^q g_{1,i}  \right)^{\sigma},
\end{eqnarray}
\begin{eqnarray}
 g'_{1,s} = z_s {\sigma \choose 2} \left( g_0 + \sum_{i=1}^q  \omega_{s}^{\delta(s,i)} g_{1,i}  \right)^{\sigma-2} \left( \sum_{j=1}^q \omega_{s}^{\delta(s,j)} g_{2,j}  \right)^{2}
\end{eqnarray}
and
\begin{eqnarray}
 g'_{2,s} = z_s \sigma \left( g_0 + \sum_{i=1}^q \omega_{s}^{\delta(s,i)} g_{1,i}  \right)^{\sigma-1} \left( \sum_{j=1}^q \omega_{s}^{\delta(s,j)} g_{2,j} \right),
\end{eqnarray}
\end{subequations}
where $s=1,\ldots,q$ and the primed (unprimed) $g$'s represent ppf's in generation $M+1$ ($M$). Since these ppf's diverge in the thermodynamic limit ($M \rightarrow \infty$), we have to work with ratios of them, which will be defined here as $R_{1,s} = \dfrac{g_{1,s}}{g_0}$ and $R_{2,s} = \dfrac{g_{2,s}}{g_0}$. We are thus left with $2q$ recursion relations (RRs) for these ratios, being
\begin{subequations}
\label{eqRRs}
\begin{eqnarray}
 R'_{1,s} =  z_s {\sigma \choose 2}\frac{ \left(1 + \sum_{k=1}^q  \omega_{s}^{\delta(s,k)} R_{1,k}\right)^{\sigma-2} \left( \sum_{i=1}^q \omega_{s}^{\delta(s,i)} R_{2,i}  \right)^{2}}{ \left( 1 + \sum_{j=1}^q R_{1,j}  \right)^{\sigma}}
\end{eqnarray}
and
\begin{eqnarray}
 R'_{2,s} = z_s \sigma \frac{ \left(1 + \sum_{k=1}^q  \omega_{s}^{\delta(s,k)} R_{1,k}\right)^{\sigma-1} \left( \sum_{i=1}^q \omega_{s}^{\delta(s,i)} R_{2,i} \right)}{ \left( 1 + \sum_{j=1}^q R_{1,j}  \right)^{\sigma}}.
\end{eqnarray}
\end{subequations}
for $s=1,\ldots,q$. These RRs converge as $M \rightarrow \infty$ and their physical (i.e., real and non-negative) fixed points determine the thermodynamic phases of the model for given $\sigma$ and $q$. The stability of these phases is analyzed through the leading eigenvalue, $\Lambda$, of the jacobian matrix. This is a $2q \times 2q$ matrix with entries $J_{\vec{a},\vec{b}}= \dfrac{\partial R_{\vec{a}}}{\partial R_{\vec{b}}}$, where $\vec{a} \equiv ( k, s)$ and $\vec{b} \equiv (k',s')$, for $k,k'=1,2$ and $s,s'=1,\ldots,q$. A given phase is stable (unstable) in regions of the parameter space where $|\Lambda| \le 1$ ($|\Lambda| > 1$). The condition $|\Lambda| = 1$ determines the spinodal of the phase. 

Similarly to the RRs for the ppf's, we may obtain the partition function of the system on the BL by attaching four subtrees with $M \rightarrow \infty$ generations to a central site. This gives
\begin{equation}
 Y^{(P)} = g_0^{\sigma+1} y^{(P)}, 
\end{equation}
for the Potts-like polymers, where
\begin{eqnarray}
 \label{eqY} 
 y^{(P)} &=& \left( 1 + \sum_{k=1}^q R_{1,k}  \right)^{\sigma+1} + {\sigma+1 \choose 2} \times \\ \nonumber
 && \sum_{s=1}^q z_s \left( 1 + \sum_{j=1}^q \omega_{s}^{\delta(s,j)} R_{1,j}  \right)^{\sigma-1} \left( \sum_{i=1}^q \omega_{s}^{\delta(s,i)} R_{2,i}  \right)^{2}.
\end{eqnarray}

The solution of the Ising-like model follows the same lines above, with the recursion relations for the ppf's and their ratios being given by expressions analogous to Eqs. \ref{eqPPFs} and \ref{eqRRs}, respectively, for $q=2$, with $\omega_s^{\delta(s,k)}$ replaced by $\omega^{2\delta(s,k)-1}$. In the same fashion, the partition function in the Ising case is $Y^{(I)} = g_0^{\sigma+1} y^{(I)}$, where $y^{(I)}$ is given by Eq. \ref{eqY} with $q=2$ and $\omega_s^{\delta(s,k)} \rightarrow \omega^{2\delta(s,k)-1}$. It is worth noticing that the existence of vacant sites in the polymeric systems breaks the equivalence between the Ising and the $q=2$ Potts models, which is the reason for discussing both solutions here.

Once one has the partition function $y^{(X)}$ at hand, for $X = P$ or $I$, the density of monomers with spin $s$ in a given phase is given by:
\begin{equation}
\left. \rho_s = \frac{z_s}{y^{(X)}} \frac{\partial y^{(X)}}{\partial z_s}\right|_{z_s = z},
\end{equation}
from which the total density $\rho=\rho_1+\cdots+\rho_q$, and the fraction $n_s=\rho_s/\rho$ of monomers with spin $s$ can be calculated. Then, the order parameter for the paramagnetic-to-ferromagnetic transition can be defined as:
\begin{equation}
m = \frac{q \max(n_1,n_2,\cdots,n_q) -1}{q-1}.
\end{equation}

As demonstrated in several works (see, e.g., Refs. \cite{Gujrati,*Oliveira09,*Oliveira10}), the reduced bulk free energy per site on the BL is
\begin{equation}
 \phi_b = - \frac{1}{2} \ln\left( \frac{Y_{M+1}}{Y_M^{\sigma}} \right),
\end{equation}
for $M \rightarrow \infty$. For the systems considered here, this yields
\begin{equation}
 \phi_b = - \frac{1}{2} \left[ \sigma (\sigma+1) \ln\left( 1 + \sum_{j=1}^q R_{1,j} \right) - (\sigma - 1)\ln y \right].
\end{equation}
Whenever two phases (let us say $A$ and $B$) coexist, their coexistence points, lines or surfaces are determined by the condition $\phi_b^{(A)}=\phi_b^{(B)}$.

As usually observed in the grand-canonical analysis of polymer models, a non-polymerized (NP) phase exists in magnetic polymers for small fugacities $z$. In fact, a simple inspection of Eqs. \ref{eqRRs} makes it clear that these RRs have a fixed point solution where $R_{k,s}=0$ for all $k=1,2$ and $s=1,\ldots,q$. This indeed corresponds to the NP phase, once this yields $\rho_1^{(NP)}= \cdots = \rho_q^{(NP)}=0$ and $\phi^{(NP)}=0$. Thanks to the simplicity of this fixed point, it is easy to determine its stability region for the Ising-like ($I$) and standard Potts-like ($P$) models, being
\begin{equation}
z^{(I)}\leq\frac{\omega}{\sigma(\omega^2+1)} \quad \text{and} \quad  z^{(P)} \leq\frac{1}{\sigma (\omega+q-1)},
\label{eqSpinNP2}
\end{equation} 
respectively, where the equalities define the spinodals of the NP phase in each case. 

For large $z$ and/or $\omega$ the systems are polymerized and, because of their magnetic character, one finds two different polymerized phases: a disordered paramagnetic (PP) and an ordered ferromagnetic (FP) phase. The former one is characterized by a fixed point solution of type $R_{i,1}=R_{i,2}=\cdots=R_{i,q}$, for $i=1,2$, such that $n_1^{(PP)}= \cdots = n_q^{(PP)}=1/q$ and, then, $m=0$. In the FP phase a symmetry breaking is observed in the RRs, with one of the spin states (let us say, $s=1$) dominating over the others, such that $R_{i,1}>R_{i,2}=\cdots=R_{i,q}$. This leads to $n_1^{(FP)} > n_2^{(FP)} = \cdots = n_q^{(FP)}$, yielding $m>0$ in the FP phase.

Since Eqs. \ref{eqRRs} reduce to only two RRs in the PP phase, it is straightforward to obtain their solutions using a software such as the Mathematica, but the expressions are too large to be given here. For general $\sigma$ and $q$, we find a critical line separating this phase from the NP one, for small values of $\omega$, as discussed in more detail in the following sections. This critical line ends at a tricritical point, which for the standard Potts-like polymers is located at [see Appendix \ref{secApendTP}]
\begin{equation}
z_{\theta}=\frac{\sigma-1}{q\sigma^2}  \quad \text{and} \quad \omega_{\theta}=\frac{\sigma+q-1}{\sigma-1}.
\label{eqThetaPointPotts}
\end{equation}
This point can be identified as the $\theta$ point, where the coil-globule transition takes place in classical models for $\theta$-polymers (as, e.g., the ISAW model), once the PP phase has the same characteristics of the polymerized phase in such models. For the Ising-like polymers, we obtain
\begin{equation}
z_{\theta}=\frac{\sigma-1}{2\sigma^2}  \quad \text{and} \quad    \omega_\theta=\frac{\sigma+\sqrt{2\sigma-1}}{\sigma-1},
\label{eqThetaPointIsing}
\end{equation}
where $z_{\theta}$ is identical to the one for the $q=2$ Potts in Eq. \ref{eqThetaPointPotts}. The results for $\omega_{\theta}$ (and those for the spinodals of the NP phase in Eq. \ref{eqSpinNP2}), on the other hand, make it clear that the Ising-like polymers are not related to the $q=2$ Potts-like model through a simple redefinition of $\omega$. As it will be demonstrated in Sec. \ref{secResq2}, this leads to quite different critical properties for these two-state models.

\section{Results for $q=1$ Potts-like polymers (the non-magnetic case)}
\label{secResq1}

For $q=1$ all monomers are identical, so that the system is non-magnetic. Nevertheless, it is instructive to start our analyses with this case for two main reasons: first, to remind the BL solution of the classical ISAW model \cite{Orr,Wall,Domb}; and second, to verify how it is affected by the interactions between NN monomers consecutive along the chain. In fact, in contrast with the magnetic polymers, in the ISAW model (and other typical models for $\theta$ polymers as well) no interaction is considered between bonded monomers.

\begin{figure}[!t]
 \includegraphics[width=8.cm]{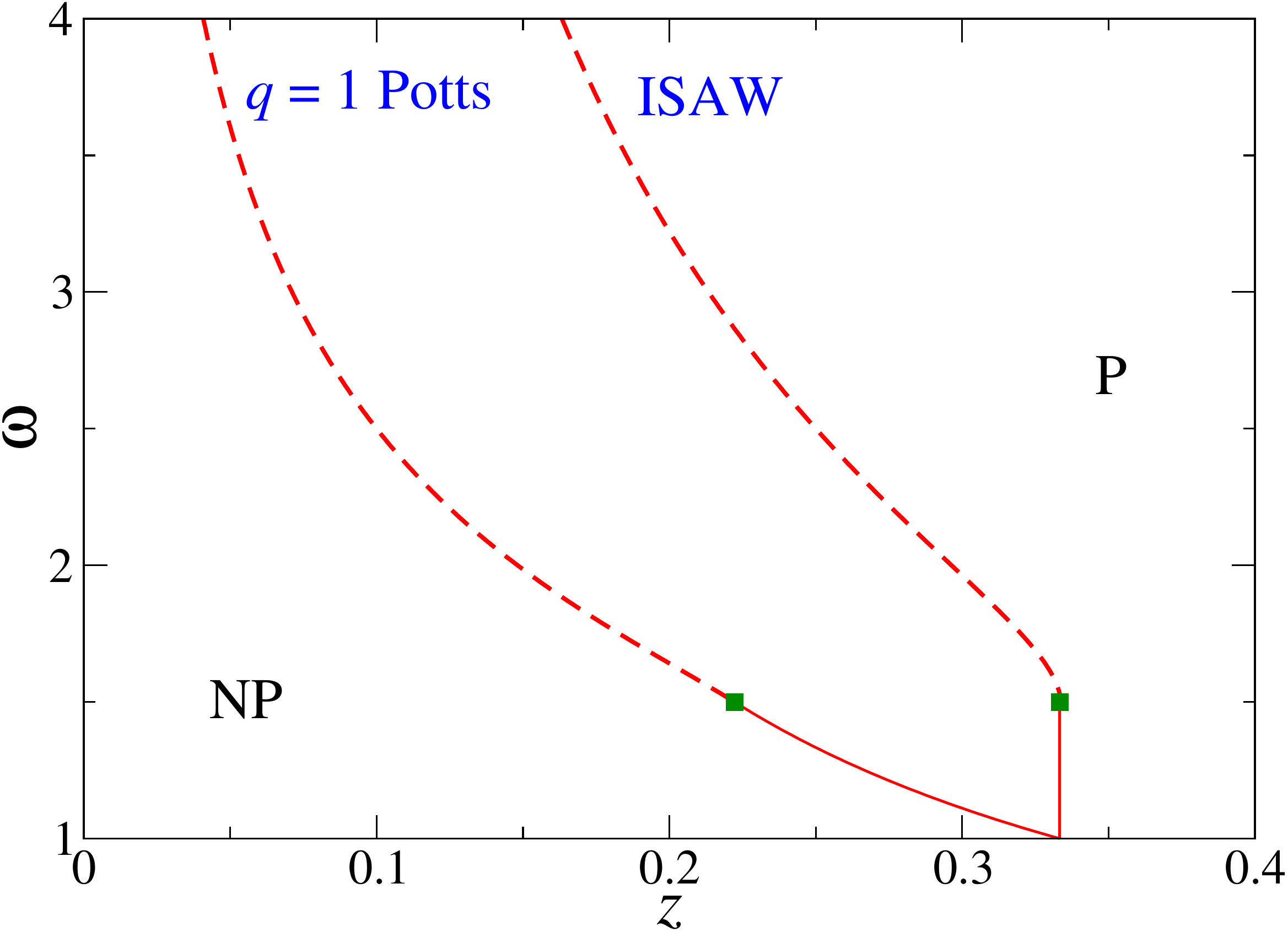}
 \caption{Phase diagrams for the ISAW (right) and the $q=1$ Potts-like polymers (left lines). The continuous and discontinuous transition lines, separating the NP and P phases, are denoted by solid and dashed lines, respectively. The square dots represent the $\theta$-points. These results are for a BL with $\sigma=3$, and similar ones are obtained for larger $\sigma$.}
 \label{fig4}
\end{figure}

Note that Eqs. \ref{eqRRs} reduce to only two RRs when $q=1$ and the PP and FP phases become a single polymerized (P) phase in this case. Therefore, this system depends on only two thermodynamic parameters $z=z_1$ and $\omega=\omega_1$, besides $\sigma$. The BL solution for the ISAW model is obtained by making $\omega_1^{\delta(s,i)} R_{2,1} \rightarrow R_{2,1}$ in Eqs. \ref{eqRRs} for $q=1$ and, similarly to the Potts-like system, it also displays a NP phase, where $R_{1,1}=R_{2,1}=0$, and a polymerized (P) phase, with $R_{1,1}>0$ and $R_{2,1}>0$ \cite{Serra,Botelho,Lise}. The NP phase is stable for $z \le 1/\sigma$, such that its spinodal is independent of $\omega$ in the ISAW. This stability limit coincides with the one for the P phase for $\omega \le \sigma/(\sigma-1)$, giving rise to a critical P-NP transition line at $z_c=1/\sigma$. However, for $\omega>\sigma/(\sigma-1)$ there is a region of coexistence between these phases and they are separated by a first-order transition line that meets the critical one at the $\theta$ point: $z_{\theta} = 1/\sigma$ and $\omega_{\theta} = \sigma/(\sigma-1)$ \cite{Serra,Botelho,Lise}. These results are summarized in Fig. \ref{fig4} for a BL with $\sigma = 3$. 

For the $q=1$ Potts-like polymers, the interaction between bounded monomers yields a considerable quantitative change in the BL solution, though the phase diagram is still qualitatively the same as for the ISAW model. For instance, now the NP phase is stable in a more restricted region: $z \le 1/(\omega \sigma)$ (see Eq. \ref{eqSpinNP2}), once its spinodal becomes $\omega$-dependent. This obviously leads to a change in the $z$-coordinate of the $\theta$-point, which is now $z_{\theta} = (\sigma-1)/\sigma^2$, but this point is located at the same $\omega$ as in the ISAW (see Eq. \ref{eqThetaPointPotts}). This diagram, for $\sigma=3$, is also presented in Fig. \ref{fig4}, where it is compared with the one for the ISAW model.

\section{Results for Ising-like and $q=2$ Potts-like polymers}
\label{secResq2}

Let us now analyze the simplest magnetic case, where each monomer $i$ can be found in only two states ($s_i=1,2$), as recently considered for the Ising-like model on the square and cubic lattices in Refs. \cite{Garel,Foster,Faizullina}. Following these works, we will investigate here only the standard case, where $\omega_1 = \omega_2 = \omega$. Therefore, we are left again with two thermodynamic variables: $z=z_1=z_2$ and $\omega$.

As discussed in Sec. \ref{secModel}, in this case the magnetic polymers present three stable phases: a non-polymerized (NP), a paramagnetic polymerized (PP) and a ferromagnetic polymerized (FP) phase. For small $\omega$ the spinodal of the PP phase coincides with the one for the NP phase (Eq. \ref{eqSpinNP2}), giving rise to critical NP-PP transition lines. These lines are expected to end at the $\theta$-points (displayed in Eqs. \ref{eqThetaPointPotts} and \ref{eqThetaPointIsing}, for the Potts- and Ising-like models, respectively), above which a first-order transition line is expected between the NP and PP phases. 

\begin{figure}[!t]
 \includegraphics[width=8.cm]{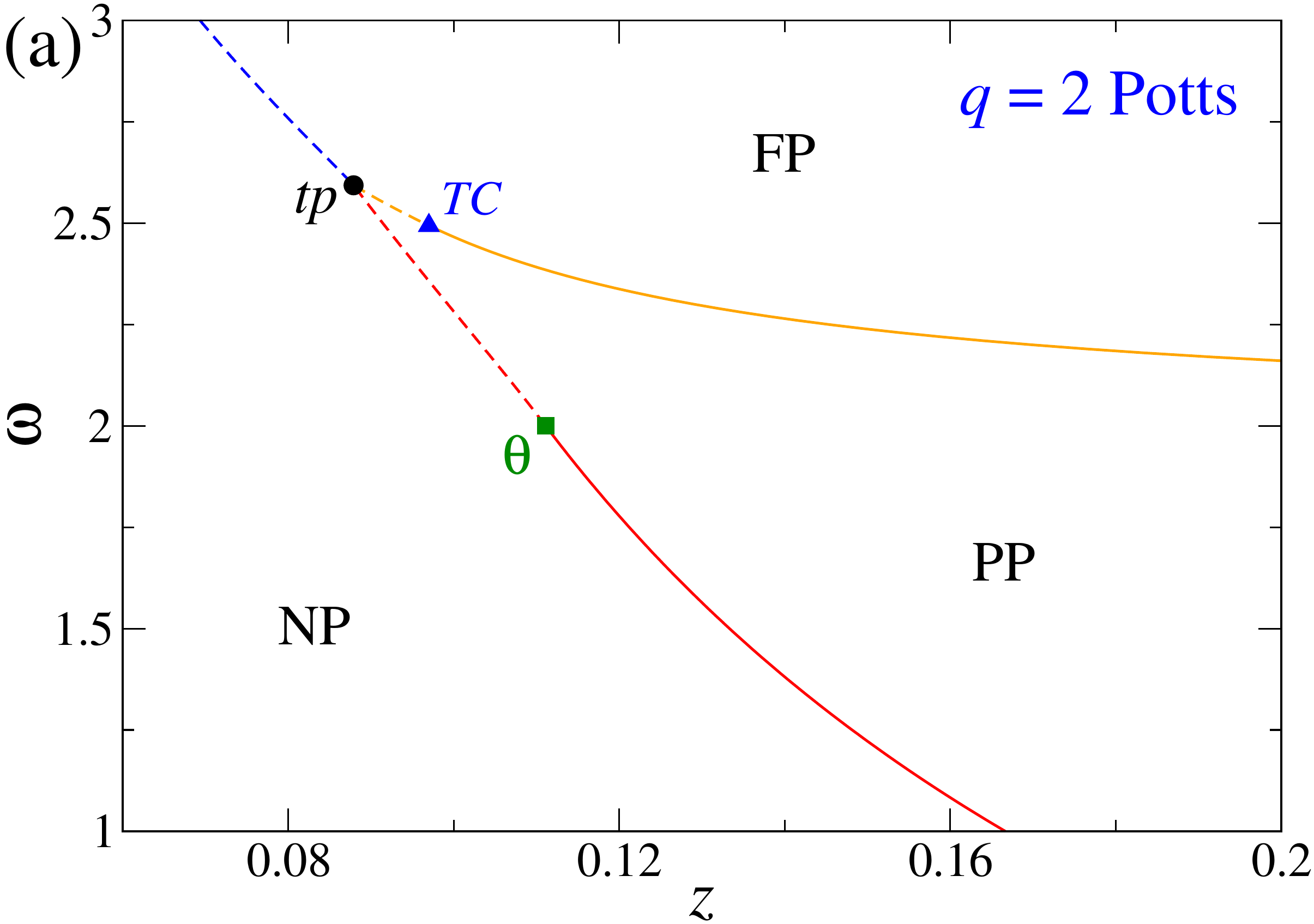}
 \includegraphics[width=8.cm]{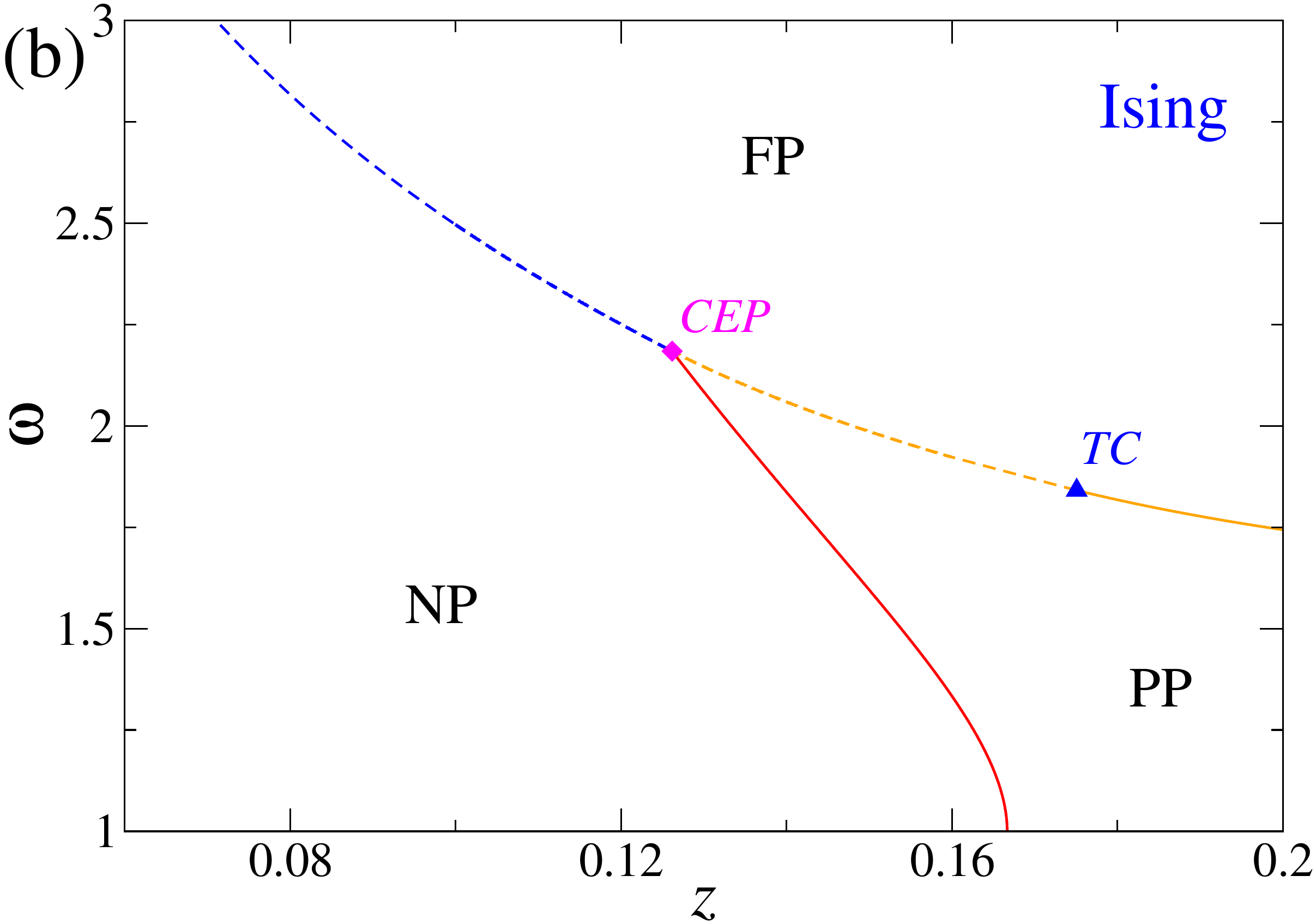}
 \includegraphics[width=8.cm]{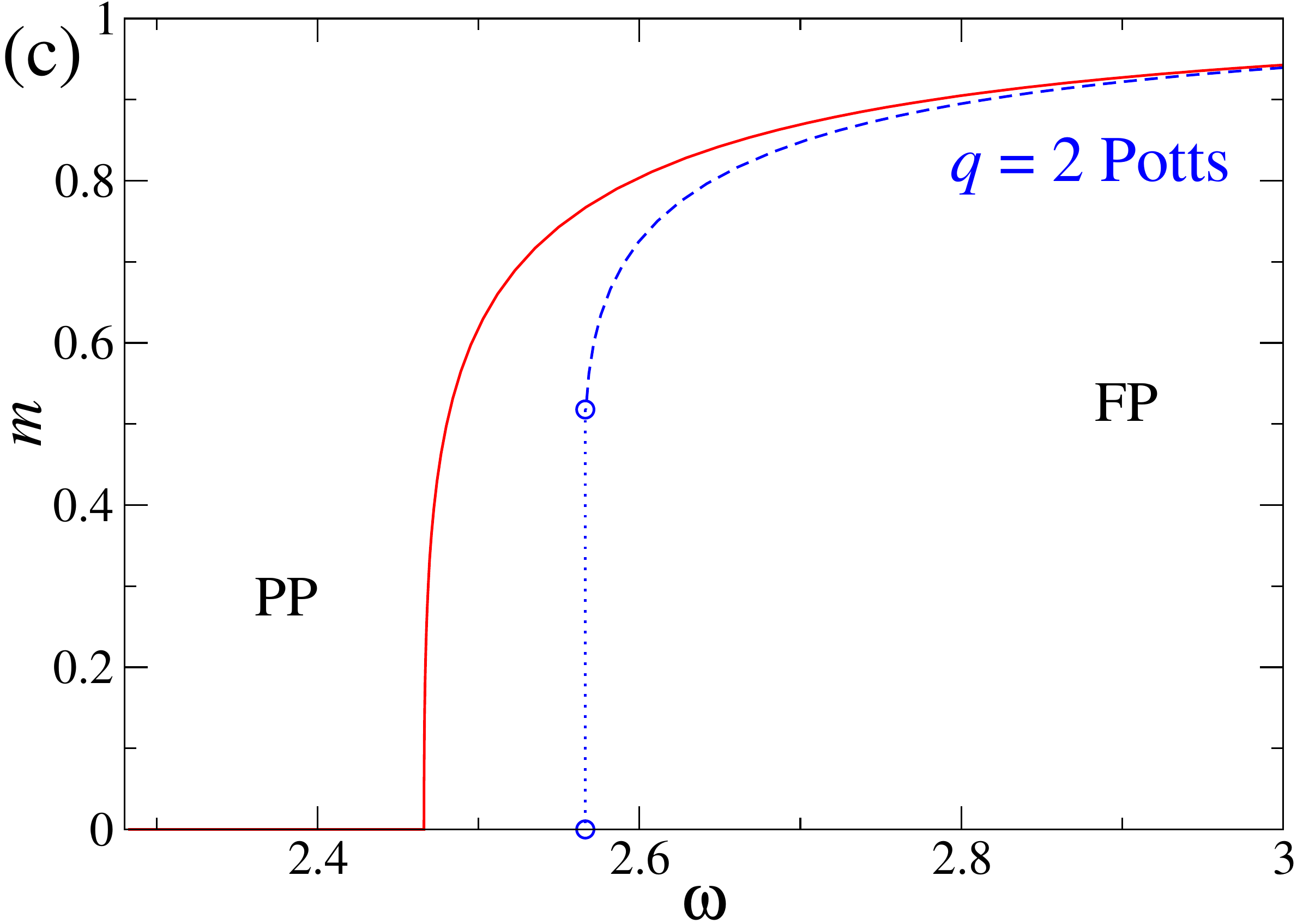}
\caption{Phase diagrams for (a) standard $q=2$ Potts-like and (b) Ising-like magnetic polymers on a BL of ramification $\sigma=3$. The solid and dashed lines denote continuous and discontinuous transitions lines, respectively. The triple (tp) point, the critical-end-point (CEP), as well as the tricritical (TC and $\theta$) points are all indicated. (c) Magnetization $m$ versus $\omega$ for the $q=2$ Potts system, for $z=0.09$ (blue dashed) and $z=0.10$ (solid red line), where the discontinuous and continuous PP-FP transition lines are crossed, respectively.}
 \label{fig5}
\end{figure}

This is exactly the scenario found for the $q=2$ Potts model, as demonstrated in the phase diagram of Fig. \ref{fig5}(a), for $\sigma=3$. In this diagram, we find also a second tricritical point related to the order-disorder transition (of spins) between the PP and FP phases. Namely, for large $z$ these polymerized phases are separated by a critical line that ends at a tricritical (TC) point, below which the PP-FP transition becomes discontinuous. These behaviors are confirmed in Fig. \ref{fig5}(c), where the magnetization $m$ is shown as a function of $\omega$, near the PP-FP transition region, for a value of $z$ below and another one above the TC point. The coordinates of this TC point, for $3 \le \sigma \le 5$, are displayed in Tab. \ref{tab1}, where one sees that both $z_{TC}$ and $\omega_{TC}$ decreases as the lattice coordination augments. For large $\omega$, we find also a NP-FP coexistence in the system and the associated first-order transition line meets the NP-PP and PP-FP ones at a triple (tp) point [see Fig. \ref{fig5}(a)]. The loci of the triple points, for $3 \le \sigma \le 5$, are summarized in Tab. \ref{tab2}. Importantly, these results reveal that the polymer collapse transition and the magnetic transition do not happen together in the $q=2$ Potts-like polymers.

\begin{table*}[t] \centering
\caption{Coordinates ($z_{TC},\omega_{TC}$) of the tricritical points for the PP-FP transition in the magnetic polymers and ($\bar{z}_{TC},\bar{\omega}_{TC}$) for the paramagnetic-to-ferromagnetic transition in the dilute models. In both cases, the results are for the standard $q=2$ Potts ($P$) and the Ising ($I$) model defined on BLs of ramification $\sigma$. The loci ($z_{CEP},\omega_{CEP}$) of the critical-end-points found for the Ising-like polymers are also shown.}
\begin{ruledtabular}
\begin{tabular}{lllllllllll}
  $\sigma$   &   $z_{TC}^{(P)}$   &   $\omega_{TC}^{(P)}$   &   $\bar{z}_{TC}^{(P)}$   &   $\bar{\omega}_{TC}^{(P)}$   &   $z_{TC}^{(I)}$   &   $\omega_{TC}^{(I)}$   &   $\bar{z}_{TC}^{(I)}$   &    $\bar{\omega}_{TC}^{(I)}$   &   $z_{CEP}$   &    $\omega_{CEP}$   \\
  \hline
  $3$  &  $0.09764$  &  $2.48794$  &  $0.11111$  &  $3.33333$  &  $0.17507$  & $1.84095$  &  $0.12647$  &  $2.97108$  &  $0.12617$  &  $2.18421$         \\ 
  $4$  &  $0.08756$  &  $1.98476$  &  $0.14238$  &  $2.40742$  &  $0.15460$  &  $1.57530$  &  $0.15597$  &  $2.19948$  &  $0.10428$  &  $1.84225$         \\ 
  $5$  &  $0.08227$  &  $1.71431$  &  $0.16383$  &  $2.00000$  &  $0.13769$  &  $1.43526$  &  $0.17433$  &  $1.85890$  &  $0.08840$  &  $1.65803$         \\ 
\end{tabular}
\end{ruledtabular}
\label{tab1}
\end{table*}

Figure \ref{fig5}(b) shows the phase diagram for the Ising-like polymers, for $\sigma=3$. Similarly to the $q=2$ Potts case, the NP-FP transition is discontinuous, and the PP-FP phases are separated by a critical (for large $z$) and a first-order (for small $z$) transition line, meeting at a TC point. The coordinates of these TC points are also depicted in Tab. \ref{tab1}, where one may observe that, for a given $\sigma$, $z_{TC}$ ($\omega_{TC}$) is larger (smaller) in the Ising case, i.e., $z_{TC}^{(I)}> z_{TC}^{(P)}$ and $\omega_{TC}^{(I)} < \omega_{TC}^{(P)}$. The main difference between these models is however in the $\omega$ values for $\theta$ points, once $\omega_{\theta}^{(I)}$ is considerably larger than $\omega_{\theta}^{(P)}$ (see Eqs. \ref{eqThetaPointPotts} and \ref{eqThetaPointIsing}). For example, $\omega_{\theta}^{(P)} = 2$, while $\omega_{\theta}^{(I)} \approx 2.618$ for $\sigma=3$. This is indeed expected, once in the Ising-like case there exists a repulsion between NN monomers with different spins, so that a larger $\omega$ is required to yield a $\theta$ collapse. As a consequence of this, the $\theta$ point becomes metastable in the Ising-like system, with the stable part of the critical NP-PP line ending at a critical-end-point (CEP), where it meets the NP-FP and PP-FP coexistence lines [see Fig. \ref{fig5}(b)]. 

As will be discussed in more detail in Sec. \ref{secConc}, this means that the polymer collapse and magnetic ordering transition happen together at the CEP. We notice also that while the NP and PP phases have vanishing density and magnetization at the CEP, these quantities are non-null for the FP phase there. For instance, $\rho_{CEP} = 0.78675$ and $m_{CEP} = 0.97180$ for $\sigma =3$ and $\rho_{CEP} = 0.81759$ and $m_{CEP} = 0.97802$ for $\sigma =5$ in the FP phase. Thereby, in a canonical analysis of this system, these quantities shall present a discontinuity at the collapse/magnetic transition, in consonance with the simulation results for the Ising-like polymers on the cubic lattice \cite{Garel,Foster}.

To understand the tricritical scenario above for the magnetic transitions (between the PP and FP phases), let us start noticing that in the $z \rightarrow \infty$ limit, when the lattice is fully occupied by monomers, the polymeric character of the system is completely lost. Namely, one recovers the usual (or pure) Ising and Potts models, with spins on all lattice sites, whose solutions on the BL display a critical point at $\omega_c^{(P)}=(\sigma+1)/(\sigma-1)$ for the $q=2$ Potts and at $\omega_c^{(I)}=\sqrt{\omega_c^{(P)}}$ in the Ising case (see, e.g., Ref. \cite{BaxterBook} and also the Appendix \ref{secApPottsDil}). By decreasing $z$, vacancies start appearing in the system and the magnetic polymers become similar to the annealed site-dilute magnetic models --- i.e., Ising and Potts lattice gases. As discussed in detail, e.g., by Qian \textit{et al.} \cite{Qian}, the dilute $q=2$ Potts model is closely related to the spin-1 Blume-Capel \cite{Blume,*Capel} model, where the crystal field, $\Delta$, acts as the chemical potential of the vacancies. Therefore, for small $\Delta$ (corresponding to large $z$ in our system) a critical line is expected, which ends at a tricritical point. 

Since we are not aware of studies of the annealed dilute Ising or Potts models on the BL, to confirm the reasoning above, we have solved these models in Appendix \ref{secApPottsDil}, associating a fugacity $\bar{z}$ to each site occupied by a magnetic ion and a weight $\bar{\omega}$ to each pair of NN sites with identical spins. (In the Ising case, NN sites with different spins also have a weight $\bar{\omega}^{-1}$.) The resulting phase diagrams are depicted in Fig. \ref{figAP}, where the disordered and ordered phases are indeed separated by a critical line that meets a first-order transition line at tricritical points for both the Ising and $q=2$ Potts models. The coordinates ($\bar{z}_{TC}, \bar{\omega}_{TC}$) of these points are also displayed in Tab. \ref{tab1}, for $3 \le \sigma \le 5$. Although $\bar{\omega}_{TC}^{(P)} > \bar{\omega}_{TC}^{(I)}$, as expected, we do not have a simple relation between these values, once the mapping $\bar{\omega}^{(I)} \rightarrow \sqrt{ \bar{\omega}^{(P)} }$ does not work in the dilute case. Note also that $\bar{\omega}_{TC}^{(P)}$ is considerably larger than $\omega_{TC}^{(P)}$, though their difference decreases as $\sigma$ increases. A similar behavior is observed also in the Ising case. Such differences are indeed expected, once for small fugacities the typical spatial configurations of the magnetic polymers and the site-dilute models shall become considerably different. For instance, in the polymers each monomer necessarily has at least two NN ones, while lone monomers can be found in the lattice gases. This certainly explains why $\omega_{TC} < \bar{\omega}_{TC}$, i.e., why the ordering occurs for a smaller interaction energy and/or a higher temperature in the polymeric system.

\section{Results for Potts-like polymers with $q \ge 3$}
\label{secResqmi3}

Next, we investigate $q$-state Potts-like polymers, for $3 \le q \le 10$. Once again, we will focus on the standard model, analyzing the phase behavior in the $z-\omega$ space.

\begin{figure}[t]
 \includegraphics[width=8.cm]{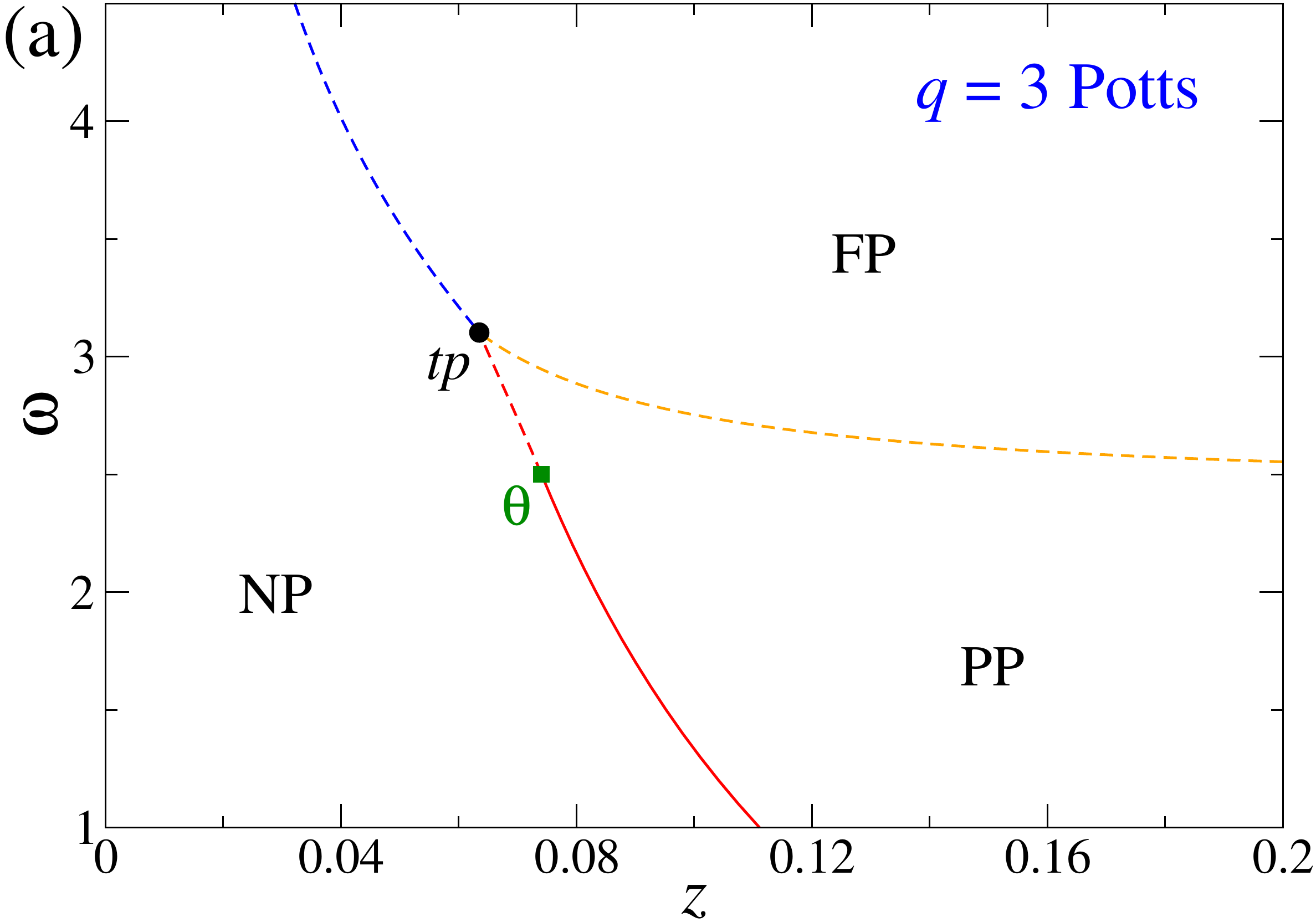}
 \includegraphics[width=8.cm]{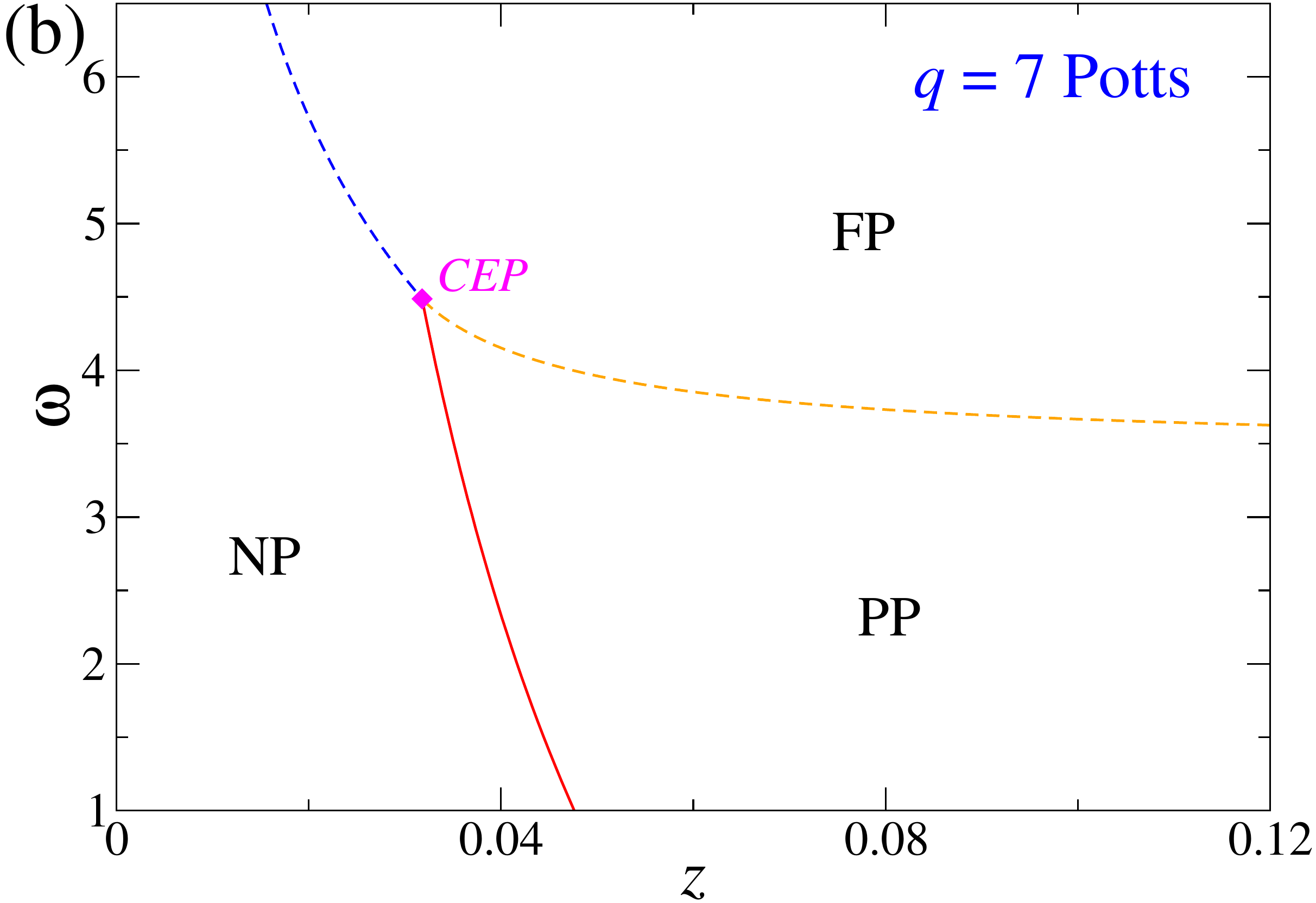}
 \caption{Phase diagrams for standard Potts polymers with (a) $q=3$ and (b) $q=7$, on a BL of ramification $\sigma=3$. The solid and dashed lines represent continuous and discontinuous transition lines, respectively. The triple (tp) point, the $\theta$ point and the critical-end-point (CEP) are all indicated.}
 \label{fig6}
\end{figure}

Let us start recalling that mean-field approaches for the (pure) standard Potts model furnish a discontinuous order-disorder transition already for $q \ge 3$ \cite{Wu}, and this is indeed the case in the BL solution of this model \cite{PottsBL2,PottsBL3}. As demonstrated in Appendix \ref{secApPottsDil}, by diluting the model (i.e., by decreasing $\bar{z}$) a first-order transition line is found separating the ordered and disordered phases, for all $q\ge 3$. Therefore, following the discussion from the previous section, a PP-FP first-order transition line is expected also in the magnetic polymers. This is indeed confirmed in Figs. \ref{fig6}(a) and \ref{fig6}(b), which respectively show the phase diagrams for $q=3$ and $q=7$, for $\sigma=3$.

Similarly to the behavior for $q=2$ [displayed in Fig. \ref{fig5}(a)], in the phase diagram of Fig. \ref{fig6}(a), for $q=3$, one sees that discontinuous PP-FP, NP-PP and NP-FP transition lines meet at a triple point. Moreover, the critical and discontinuous NP-PP lines are still connecting at a stable $\theta$ point. The very same scenario of $q=3$ is found for all $q \le 6$. The numerically estimated coordinates ($z_{tp},\omega_{tp}$) of the triple points are given in Tab. \ref{tab2}, where they are compared with the numerical values for the exact results for the $\theta$ points, given by Eq. \ref{eqThetaPointPotts}. Since $\omega_{tp} > \omega_{\theta}$, these results are demonstrating that by increasing the coupling energy $\varepsilon$ or decreasing the temperature $T$ the polymer first collapse (at the $\theta$ point) and then, in a different point (i.e., at the triple point), it undergoes the magnetic transition, for all $q\le 6$.

\begin{table}[t] \centering
\caption{Triple (tp) points for $q$-state Potts-like polymers on BLs of ramification $\sigma$. The numerical values for the exact results for the $\theta$ points (Eq. \ref{eqThetaPointPotts}) are also shown, for comparison.}
\begin{tabularx}{\columnwidth}{p{0.3cm} >{\centering\arraybackslash}X >{\centering\arraybackslash}X >{\centering\arraybackslash}X >{\centering\arraybackslash}X >{\centering\arraybackslash}X}
 \hline
 \hline
  $q$   &   $z_{tp}$   &   $\omega_{tp}$   &   $z_{\theta}$   &    $\omega_{\theta}$   \\
  \hline
                         \multicolumn{5}{c}{$\sigma=3$}                                             \\
  \hline
  $2$        &   $0.0879$      &   $2.59345$           &    $0.111111$     &    $2.00000$           \\ 
  $3$        &   $0.0635$      &   $3.10228$           &    $0.074074$     &    $2.50000$           \\ 
  $4$        &   $0.0505$      &   $3.50965$           &    $0.055555$     &    $3.00000$           \\ 
  $5$        &   $0.0421$      &   $3.86570$           &    $0.044444$     &    $3.50000$           \\ 
  $6$        &   $0.0362$      &   $4.18830$           &    $0.037037$     &    $4.00000$           \\ 
 \hline
                         \multicolumn{5}{c}{$\sigma=4$}                                             \\
 \hline
  $2$        &   $0.07645$      &   $2.06779$           &    $0.093750$     &    $1.66666$           \\ 
  $3$        &   $0.05513$      &   $2.39359$           &    $0.062500$     &    $2.00000$           \\ 
  $4$        &   $0.04367$      &   $2.65018$           &    $0.046875$     &    $2.33333$           \\ 
  $5$        &   $0.03624$      &   $2.87157$           &    $0.037500$     &    $2.66666$           \\  
  $6$        &   $0.03096$      &   $3.07008$           &    $0.031250$     &    $3.00000$           \\ 
  \hline
                         \multicolumn{5}{c}{$\sigma=5$}                                             \\
  \hline
  $2$        &   $0.06650$      &   $1.80330$           &    $0.080000$     &    $1.50000$           \\ 
  $3$        &   $0.04788$      &   $2.04056$           &    $0.053333$     &    $1.75000$           \\ 
  $4$        &   $0.03779$      &   $2.22532$           &    $0.040000$     &    $2.00000$           \\ 
  $5$        &   $0.03123$      &   $2.38326$           &    $0.032000$     &    $2.25000$           \\ 
  $6$        &   $0.02658$      &   $2.52376$           &    $0.026666$     &    $2.50000$           \\ 
 \hline
 \hline
\end{tabularx}
\label{tab2}
\end{table}

As it is clearly observed in Tab. \ref{tab2}, both $z_{tp}$ and $z_{\theta}$ decreases, while $\omega_{tp}$ and $\omega_{\theta}$ increases, for a given $\sigma$, as $q$ grows. However, $\omega_{\theta}$ increases faster than $\omega_{tp}$, in a way that for $q \ge 7$ the $\theta$ point becomes metastable. Namely, for such large $q$'s, the discontinuous NP-FP and PP-FP lines meet at the critical NP-PP line. Hence, the stable part of this critical line ends at a critical-end-point (CEP). The resulting phase diagram for $q=7$ and $\sigma=3$ is depicted in Fig. \ref{fig6}(b), and similar ones are found for larger $q$'s and $\sigma$'s. The coordinates of the CEPs --- for $\sigma=3$ and $5$, and $q$ up to 10 --- are displayed in Tab. \ref{tab3}, where one sees that $\omega_{CEP}$ ($z_{CEP}$) is an increasing (decreasing) function of $q$. Thus, similarly to the Ising-like system, the $q \ge 7$ Potts-like polymers become both collapsed and magnetically ordered at a CEP.

\begin{table}[b] \centering
\caption{Critical-end-points (CEPs), where the critical NP-PP transition lines end, for $q$-state Potts-like magnetic polymers on BLs of ramification $\sigma$.}
\begin{tabularx}{\columnwidth}{p{0.3cm} >{\centering\arraybackslash}X >{\centering\arraybackslash}X >{\centering\arraybackslash}X >{\centering\arraybackslash}X >{\centering\arraybackslash}X}
 \hline
 \hline
             &  \multicolumn{2}{c}{$\sigma=3$}  &   \multicolumn{2}{c}{$\sigma=5$}    \\
  \hline
  $q$        &   $z_{CEP}$  &   $\omega_{CEP}$  &   $z_{CEP}$  &    $\omega_{CEP}$   \\
  \hline
  $7$        &   $0.0317$      &   $4.48690$           &    $0.02312$     &    $2.65054$            \\ 
  $8$        &   $0.2833$      &   $4.76390$           &    $0.02048$     &    $2.76497$            \\ 
  $9$        &   $0.0255$      &   $5.02160$           &    $0.01840$     &    $2.86966$            \\ 
  $10$       &   $0.0233$      &   $5.26362$           &    $0.01671$     &    $2.96650$            \\ 
 \hline
 \hline
\end{tabularx}
\label{tab3}
\end{table}

\section{Results for magnetic polymers with nonmagnetic monomers}
\label{secResNSPotts}

\subsection{``Two-state" systems}

As discussed in the Introduction, Michieletto \textit{et al.} \cite{Michieletto} have studied a model for the chromatin folding coupled to the dynamics of epigenetic marks spreading, consisting of bead-and-spring chains where each bead carries a ``color" (or mark) $s$, with $s=\{1,2,3\}$. The truncated Lennard-Jones potential (and cutoffs) considered there was such that a self-attraction exists between pairs of beads ($s_i,s_j$) whenever $s_i=s_j=1$ or $s_i=s_j=2$, with the same energy in both cases, whereas only a steric repulsion exists otherwise. 

Motivated by these works, we investigate here 3-state Potts polymers with $\omega_{1}=\omega_2=\omega$ and $\omega_3=1$, which can be seen as a lattice version for the model above \cite{Michieletto}. Note that this model corresponds to $q=2$ Potts-like polymers where nonmagnetic (i.e., ``uncolored" or unmarked) monomers are also present along the chain, having the same fugacity of the other ones.
In fact, its phase diagram (not shown) is qualitatively identical to the one in Fig. \ref{fig5}(a) for the standard $q=2$ Potts-like model. However, when the nonmagnetic monomers are present, the transition points occur for a smaller $z$ and a larger $\omega$. For $\sigma=5$, for example, the $\theta$ point is at $(z_{\theta},\omega_{\theta})=(0.05546,1.83826)$, the tricritical (TC) point in the magnetic transition at $(z_{TC},\omega_{TC})=(0.07225,1.81766)$, and the triple (tp) point at $(z_{tp},\omega_{tp})=(0.05327,1.95190)$. This scenario is in contrast with the simulation results from Refs. \cite{Michieletto}, where evidence of a simultaneous collapse and magnetic first-order transition was found.

To confirm that the inclusion of nonmagnetic monomers in two-state magnetic polymers does not alter their qualitative thermodynamic behavior, we have studied also the Ising-like model with such ``impurities". Namely, we considered again a $q=3$ system where only monomers in states $s=1,2$ may interact among them, according to the Ising coupling. The phase diagram found in this case is qualitatively the same as the one for the Ising-like polymers, as depicted in Fig. \ref{fig5}(b), and for $\sigma=5$, for example, one finds $(z_{CEP},\omega_{CEP})=(0.06789,1.84910)$ and $(z_{TC},\omega_{TC})=(0.22364,1.51052)$. So, once again, the presence of nonmagnetic monomers yields an increase in the $\omega$-coordinates of these transition points when compared with those for the standard model. 

\subsection{``One-state" systems}

\begin{figure}[!t]
 \includegraphics[width=8.cm]{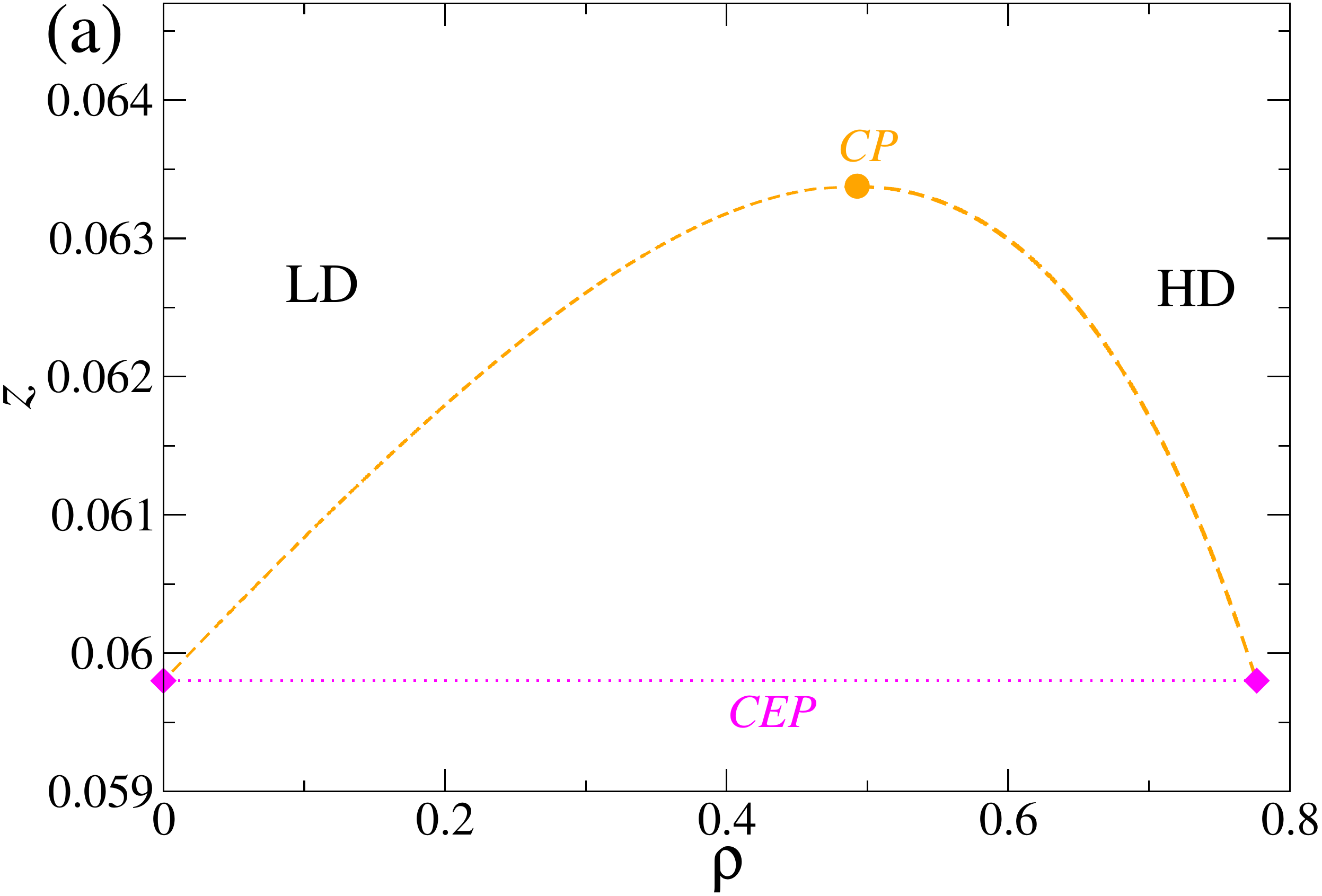}
 \includegraphics[width=8.cm]{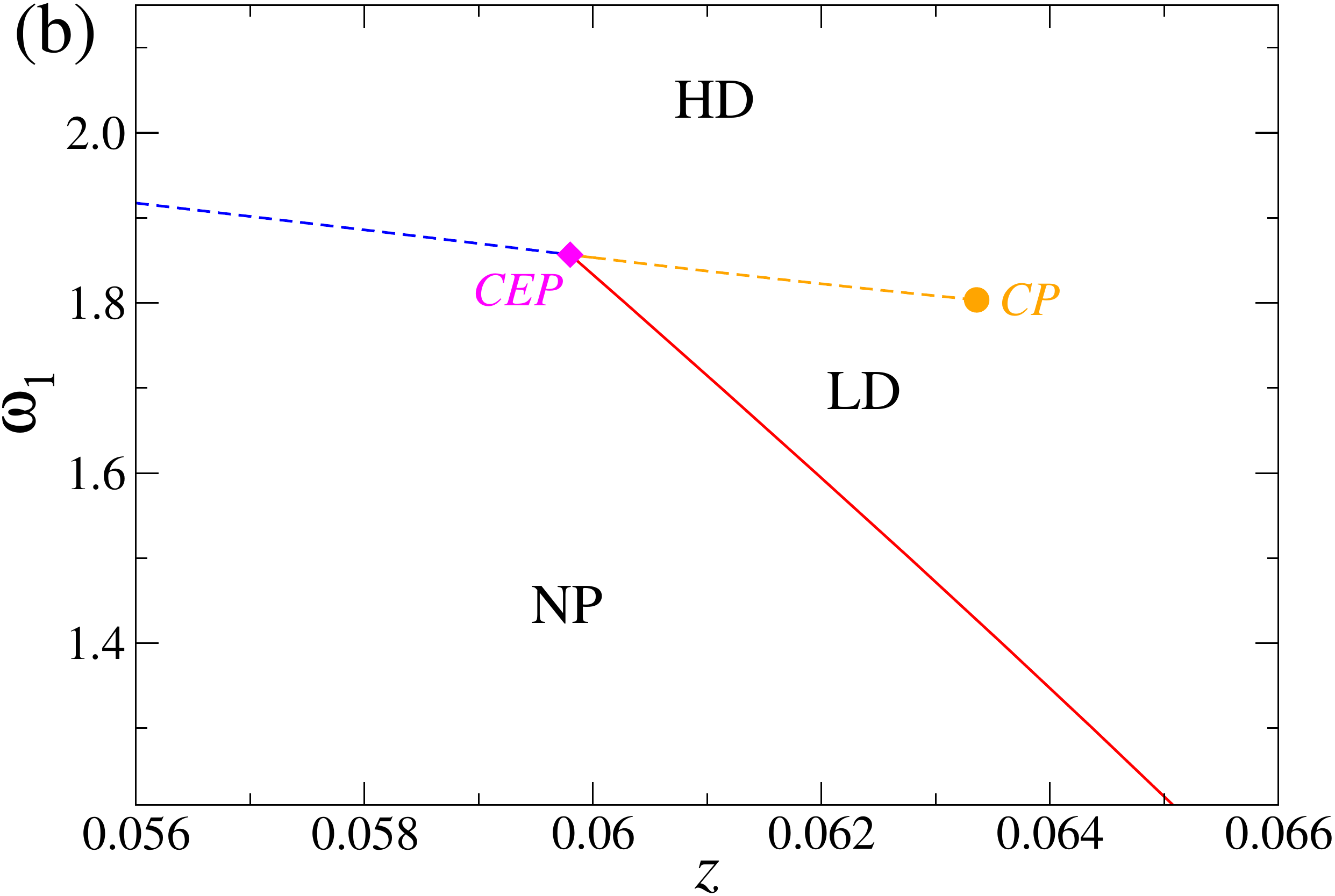}
\caption{(a) Fugacity, $z$, versus density, $\rho$, along the LD-HD coexistence line of non-standard $q=3$ Potts-like polymers with $\omega_2=\omega_3=1$ on a BL of ramification $\sigma=5$. (b) Phase diagram for the same model in variables $\omega_1$ versus $z$. The solid and dashed lines represent continuous and discontinuous transition lines, respectively. The critical point (CP) and the critical-end-point (CEP) are both indicated.}
 \label{fig7}
\end{figure}

It is interesting to analyze also situations where only monomers in a single state (let us say, $s=1$) interact with each other, provided that they are in NN sites. In this case there is no competition between two or more interacting states, which would lead to a symmetry breaking among them, yielding two stable polymerized (para- and ferromagnetic) phases in the system. In fact, as seen in Sec. \ref{secResq1}, a single polymerized phase is present in this model when nonmagnetic monomers are absent. However, by introducing such monomers in the polymer chain their competition with the interacting monomers gives rise to two types of stable polymerized phases: a high-density (HD) one, where the polymer density, $\rho$, as well as the fraction, $n_1$, of interacting monomers is large; and a low-density (LD) phase, where $\rho$ and $n_1$ are smaller. These polymerized phases coexist in a small region of the parameter space, with the coexistence line ending at a critical point (CP), above which they become indistinguishable. Figure \ref{fig7}(a) shows the values of $z$ versus $\rho$ along this coexistence line for the non-standard $q=3$ Potts-like system with $\omega_2=\omega_3=1$ and $\omega_1 \ge 1$, corresponding thus to a situation where monomers in states $s=2,3$ are nonmagnetic. This makes it clear how different are these densities not so far from the critical point, justifying the naming of the related phases. 

The complete phase diagram for this model, in space $(z,\omega_1)$, for $\sigma=5$, is depicted in Fig. \ref{fig7}(b). There, one sees that while the NP-LD transition is continuous, the NP-HD one is discontinuous, similarly to the LD-HD transition. These three transition lines meet at a CEP, where the stable part of the NP-LD critical line ends. The LD-HD coexistence line starts at the CEP and ends at a critical point (CP), as also seen in Fig. \ref{fig7}(a). Curiously, these two points are very close in the phase diagram, being located at $(z_{CEP},\omega_{CEP})=(0.05980,1.85667)$ and $(z_{CP},\omega_{CP})=(0.06337,1.80322)$ for $\sigma=5$. Hence, there exists only a very limited region of the parameter space where the LD phase is the most stable one. As an aside, we notice that phases appearing only in narrow interval of parameters have already been found in some solutions of polymers models on tree-like lattices \cite{tiago11,tiago18}, but it seems do not exist any relation between them and the LD phase.

We investigate also the $q=2$ version of the non-standard Potts-like model above, where $\omega_2=1$ and $\omega_1 \ge 1$, such that monomers in state $s=2$ are nonmagnetic. As expected, the phase diagram found for this system is qualitatively analogous to the one in Fig. \ref{fig7}(b), but now one has $(z_{CEP},\omega_{CEP})=(0.08543,1.59535)$ and $(z_{CP},\omega_{CP})=(0.08558,1.59242)$ for $\sigma=5$. Namely, the coexistence region between the LD and HD phases becomes even narrower. We remark that this system is closely related to the dynamic HP model recently studied by Faizullina \& Burovski \cite{Faizullina2}, consisting of SAWs on the square lattice composed by hydrophobic (H) and polar (P) monomers which can dynamically convert into one another. Pairs of NN monomers \textit{non-consecutive along the chain} of type P-P and H-P do not interact, whereas a self-attraction exists between the H-H ones. Hence, by identifying the state $s=1$ with H and $s=2$ with P, the only difference between the model considered here and the dynamic HP one \cite{Faizullina2} is the absence of interaction between bounded monomers in the last model. This may explain why a continuous $\theta$-like collapse transition was found in Ref. \cite{Faizullina2}, while here it occurs at a CEP, indicating that it should be discontinuous.

\section{Final discussions and conclusion}
\label{secConc}

We have presented the grand-canonical solution of several interesting models on the Bethe lattice (BL), including the standard ferromagnetic Ising-like and Potts-like magnetic polymers, modified versions of these systems where non-interacting monomers are mixed with the interacting ones, as well as the annealed site-dilute Ising and Potts models. These systems were investigated on BLs with different coordination numbers $(\sigma+1)$, but no qualitative change was observed in their thermodynamic behavior for the $\sigma$'s analyzed.

For the dilute Ising and dilute $q=2$ Potts model, we find a critical line, for large fugacity ($\bar{z}$) of magnetic ions, and a first-order transition line, for small $\bar{z}$, separating the para- and ferromagnetic phases, both lines meeting at a triple point. This is in consonance with the behavior of these systems on regular lattices (see, e.g., Ref. \cite{Qian}). For the Potts model with $q \ge 3$, the order-disorder transition is always discontinuous on the BL. Substantially, the very same behavior is observed for the transitions between the paramagnetic polymerized (PP) and ferromagnetic polymerized (FP) phases in the magnetic polymers. In fact, when $\bar{z}$ and the monomer fugacity ($z$) in the polymers are both large --- meaning that the densities, $\rho$, of monomers or magnetic ions are also large ---, the magnetic polymers shall not differ too much from the dilute models. Since the PP-FP transitions occur for relatively large $\rho$ in the Potts-like case (typically for $\rho \gtrsim 0.6$), this explains the similarity with their counterpart in the dilute models. For small $z$, on the other hand, the magnetic polymers are found in a non-polymerized phase, which does not exist in the dilute models.

To compare our results for the magnetic polymers with those from simulation studies in the canonical ensemble, we notice that in such simulations a fixed-size polymer chain and its surrounding sites (corresponding to the polymerized phases in our grand-canonical diagrams) are placed inside, and thus coexists with, an effectively infinite empty lattice (corresponding to the NP phase here). Therefore, the canonical case is given by the NP-polymerized transitions obtained here (see, e.g., Ref. \cite{Oliveira09} for a more detailed discussion about this). This means that, along the critical NP-PP lines, one has the coil phase, once the polymer density vanishes at these lines. On the other hand, along the NP-PP and NP-FP coexistence lines the polymer densities are non-vanishing, corresponding thus to globule phases; the former (later) one with disordered (ordered) spins. 

Therefore, for the Potts-like polymers with $2 \le q \le 6$, a continuous collapse transition takes place at the $\theta$-point, between a coil and a globule phase which are both magnetically disordered. The magnetic transition occurs at a different point; namely, it happens at the triple point, which exists for a larger interaction energy ($\varepsilon$) or a smaller temperature ($T$). According to our results, this second transition, between a para- and a ferromagnetic globule phase, shall be discontinuous, once both the magnetization and the polymer density present a discontinuity at the triple point. To verify that this scenario is not a feature of the BL, we have performed some preliminary flatPERM simulations for the $q=2$ Potts-like polymers on the cubic lattice and evidence of a usual $\theta$-like collapse, at $\varepsilon/k_B T \approx 0.46$, was found. This indicates that the coil-globule and the magnetic transition do indeed not happen together. We anticipate that accessing the later transition, which would occur inside the globule phase, is a difficult task with this type of simulation. More details on such simulations, extended also for other $q$'s, will be published elsewhere.

It is quite interesting that for Ising-like and Potts-like polymers with $q \ge 7$ a very different canonical scenario is found. In these systems, there is a direct transition from a paramagnetic coil to a ferromagnetic globule phase at a critical-end-point (CEP), so that the collapse and magnetic transitions occur concomitantly there. Moreover, since both the density and the magnetization of the globule phase have non-vanishing values at the CEP, the transition is discontinuous in these cases. This is in agreement with the behavior observed in simulations of the Ising-like model on the cubic lattice \cite{Garel,Foster}. In fact, our mean-field results are expected to be consistent with those for regular lattices when their dimension is high. Since for the Ising-like model on the square lattice evidence for a continuous collapse/magnetic transition was found in Refs. \cite{Foster,Faizullina}, with $\theta$ and Ising critical exponents, it seems that in the grand-canonical ensemble one should have the critical PP-FP line ending at or very close to the $\theta$-point. Further grand-canonical analysis of this system, e.g., with Monte Carlo simulations, transfer matrix methods or generalized Husimi lattices built with square clusters \cite{Nathann21,*Nathann22}, are worth to confirm this.

We have verified also that the inclusion of nonmagnetic monomers in the ``two-state'' models does not change their thermodynamic behavior. Namely, for the $q=2$ Potts-like polymers with nonmagnetic monomers we find a $\theta$-collapse separated from a discontinuous magnetic transition, whereas in the Ising-like system with nonmagnetic monomers both transitions happen together and are discontinuous. A very different phase behavior is found for the ``one-state'' model, where nonmagnetic monomers are mixed with a single type of interacting one. In this model, which is closely related to a dynamic HP model for heteropolymers \cite{Faizullina2}, the grand-canonical diagram presents a low-density and a high-density polymerized phase, which coexists in a small region of the parameter space, with the coexistence line ending at a critical point. The canonical scenario suggested for this model is analogous to the one for the Ising-like polymers, with a discontinuous collapse/magnetic transition.

\acknowledgments

We acknowledge partial financial support from CNPq, CAPES, FAPEMIG and FAPERJ (Brazilian agencies). We thank J. F. Stilck for a critical reading of the manuscript.

\appendix

\section{Determination of the $\theta$ points}
\label{secApendTP}

To obtain the exact coordinates of the $\theta$ points for the magnetic polymers on the BL, we start recalling that in the PP phase the recursion relations in Eqs. \ref{eqRRs} reduce to only two equations, regardless of $q$, because $R_{1,1}=\cdots = R_{1,q} = R_1$ and $R_{2,1}=\cdots = R_{2,q} = R_2$. Therefore, in the fixed point of this phase, equation \ref{eqRRs}b becomes a $\sigma$\textit{nt} degree polynomial:
\begin{subequations}
\label{eqRRsPP}
\begin{eqnarray}
 P(R_1) &=& \left( 1 + q R_1  \right)^{\sigma} - x\sigma z  \left[1 + x R_1\right]^{\sigma-1} \\ \nonumber
  &=&  \left( 1-x\sigma z\right) + \sigma\left[q -(\sigma-1)x^2 z\right] R_1 + \mathcal{O}(R_1^2), 
\end{eqnarray}
where
\begin{eqnarray}
 x=\omega+q-1.
\end{eqnarray}
\end{subequations}
The real and positive roots of $P(R_1)$ define the physical solutions for $R_1$. Along the critical NP-PP line, there is only one of such a roots for the PP phase and it is equal to the NP solution there; namely, $R_1=0$ along this line. Therefore, the zeroth-order term of $P(R_1)$ vanishes in this critical condition. In fact, this yields $z=1/(\sigma x)$, which corresponds to the spinodal of the NP phase, as can be easily seen by comparing this with Eq. \ref{eqSpinNP2}.

At the tricritical condition the solutions become triply degenerated, so that two roots of $P(R_1)$ become identical to the NP one  there. This means that both the zeroth-order and the first-order terms of $P(R_1)$ vanish at the $\theta$ points \footnote{See, e.g., Ref. \cite{Serra} for a similar discussion on this point for the ISAW model.}. These terms are explicitly given in Eq. \ref{eqRRsPP}a and after a simple algebra we obtain the loci of the $\theta$ points for the Potts-like polymers, as given in Eq. \ref{eqThetaPointPotts}.

As pointed in Sec. \ref{secModel}, the recursion relations for the Ising-like polymers on the BL are given by expressions analogous to Eqs. \ref{eqRRs} with $q=2$ and $\omega_s^{\delta(s,j)}$ replaced by $\omega^{2\delta(s,j)-1}$. Hence, one still finding the polynomial \ref{eqRRsPP}a in this case, but with $x = \omega + 1/\omega$. Thereby, following the same steps as above, one readily arrives at the result in Eq. \ref{eqThetaPointIsing} for the $\theta$ points in this case.

\section{Dilute Ising and Potts models on the Bethe lattice}
\label{secApPottsDil}

\begin{figure}[!b]
 \includegraphics[width=8.cm]{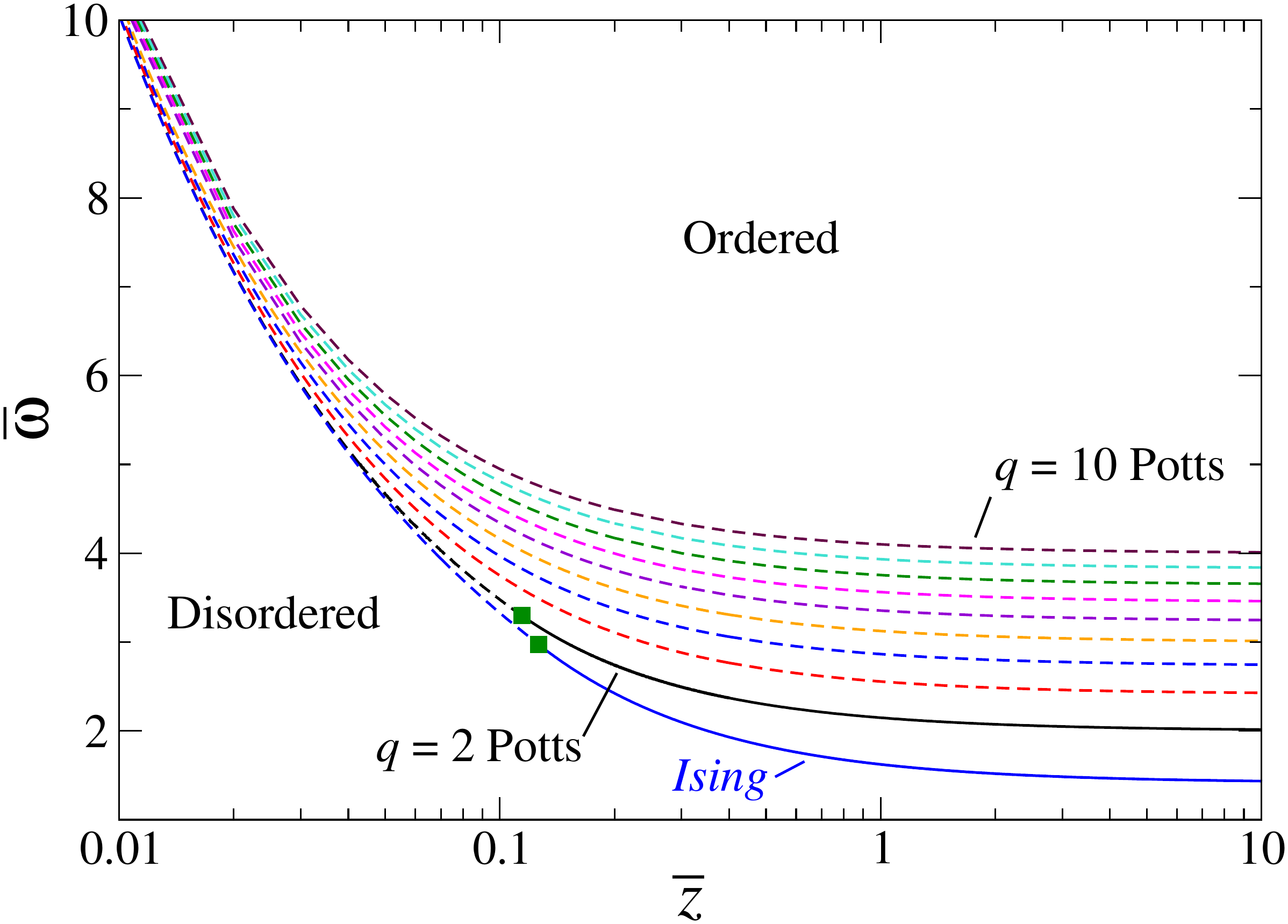}
 \caption{Phase diagrams in variables $\bar{z} - \bar{\omega}$ for the (annealed) site dilute Ising and $q$-state Potts models, for $q=2,3,..,10$, on a BL with ramification $\sigma=3$. Continuous and discontinuous transition lines are represented by solid and broken lines, respectively. The tricritical points for the Ising and $q=2$ Potts models are denoted by the green dots.}
 \label{figAP}
\end{figure}

Let us now consider the site-dilute Ising and Potts models in their annealed version --- where the vacancies are free to move and are in equilibrium with their surroundings --- on the BL, once again for ferromagnetic interactions and zero external magnetic field. As above, each site can be empty [partial partition function (ppf) $g_0$] or occupied by a particle in state $s$ [ppf $g_s$], and defining the ratios $R_s = g_s/g_0$, we obtain $q$ recursion relations (RRs) for them in the Potts case, being
\begin{equation}
 R'_{s} = \frac{\bar{z} \left( 1 + \sum_{i=1}^q  \bar{\omega}^{\delta(s,i)} R_{i}  \right)^{\sigma}}{\left( 1 + \sum_{j=1}^q R_{j}  \right)^{\sigma}},
\label{eqRRsDilute}
\end{equation}
with $s=1,\ldots,q$. Note that, similarly to the magnetic polymers considered above, we are attributing a fugacity $\bar{z}$ to each site occupied by a magnetic ion and a weight $\bar{\omega}$ to each pair of NN sites occupied with the same spin. The RRs for the dilute Ising model are given by Eqs. \ref{eqRRsDilute}, with $q=2$ and $\bar{\omega}^{\delta(s,i)} \rightarrow \bar{\omega}^{2\delta(s,i)-1}$. In contrast to the polymeric case, the RRs in Eq. \ref{eqRRsDilute}, as well as those for the Ising case do not admit an ``empty lattice'' solution [corresponding to $R_s = 0$ $\forall$ $s$], but they display the disordered [with $R_1 = \cdots = R_q$] and the ordered [e.g., with $R_1>R_2 = \cdots = R_q$] fixed points, as expected. Following the same type of calculation discussed in Sec. \ref{secModel}, adapted to the present models, we may obtain the spinodals, densities, free energies and so on for these phases. For the sake of conciseness, we will omit the related expressions here, going directly to the results.

As an aside, we notice that the RRs for the non-dilute case (i.e., for the full lattice limit) can be obtained from Eqs. \ref{eqRRsDilute} (and related equations for the Ising case) by defining $X_s = R_s/\bar{z}$ and then taken the limit $\bar{z} \rightarrow \infty$. Results for these pure models on the BL have been reported by several authors in the last decades \cite{PottsBL1,PottsBL2,PottsBL3,PottsBL4,PottsBL5}. The $q=2$ Potts model displays a continuous transition at $\bar{\omega}_c^{(P)} = (\sigma+1)/(\sigma-1)$ \cite{PottsBL2,BaxterBook}, whereas for $q \ge 3$ a coexistence region exists between the ordered and disordered phases, with a first-order transition taking place at $\bar{\omega}^* = (q-2)/[(q-1)^{(\sigma-1)/(\sigma+1)}-1]$ \cite{PottsBL2,PottsBL3}. This is consistent with the widely known fact that mean-field approaches for the Potts model yield a first-order transition for any $q \ge 3$ \cite{Wu}. In the pure Ising model the transition is continuous and located at $\bar{\omega}_c^{(I)}=\sqrt{\bar{\omega}_c^{(P)}}$.

By including vacancies in the system, we find that the critical point $\bar{\omega}_c$, for the Ising or $q=2$ Potts models, is the end point of a critical line that extends to finite $\bar{z}$. For small $\bar{z}$, however, the transition becomes discontinuous and the coexistence line meets the critical one at a tricritical point (see Fig. \ref{figAP}). The coordinates of the tricritical points for some $\sigma$ are displayed in Tab. \ref{tab1} of the text above. For the Potts model with $q \ge 3$, the coexistence found for $\bar{z} \rightarrow \infty$ continues existing for finite $\bar{z}$, giving rise to first-order transition lines, and the order-disorder transition is always discontinuous in these cases. Figure \ref{figAP} presents such transition lines for $\sigma=3$ and similar ones are found for larger $\sigma$.

\end{document}